\documentclass[11pt]{article}

\usepackage{fullpage}
\usepackage{amsmath}
\usepackage{amssymb}
\usepackage{mathrsfs}
\usepackage{setspace}
\usepackage{bbm}
\usepackage{dsfont}

\usepackage{graphics}

\usepackage[latin1]{inputenc}
\usepackage[pdftex]{graphicx}

\usepackage{color}
\usepackage[american]{babel}

\newcommand{\be}{\begin{equation}}
\newcommand{\ee}{\end{equation}}
\newcommand{\bes}{\begin{equation*}}
\newcommand{\ees}{\end{equation*}}
\newcommand{\bea}{\begin{eqnarray}}
\newcommand{\eea}{\end{eqnarray}}
\newcommand{\beas}{\begin{eqnarray*}}
\newcommand{\eeas}{\end{eqnarray*}}

\newcommand{\bb}{\mathbb}
\newcommand{\p}{\partial}

\newcommand{\irr}{{\rm irr}}

\def\a{\alpha}

\newcommand{\C}{\mathbb{C}}
\newcommand{\R}{\mathbb{R}}
\newcommand{\Z}{\mathbb{Z}}

\begin{document}
\numberwithin{equation}{section}
{
\begin{titlepage}
\begin{center}

\hfill \\
\hfill \\
\vskip 0.75in

%\

{\Large \bf Black Hole Monodromy and Conformal Field Theory}\\

\vskip 0.4in

 { Alejandra Castro${}^a$, Joshua M. Lapan${}^b$, Alexander Maloney${}^b$, and Maria J. Rodriguez${}^a$}\\

\vskip 0.3in

{\it$^a$ Center for the Fundamental Laws of Nature, Harvard University, Cambridge, MA 02138, USA }  \vskip .5mm              
{\it $^b$ Physics Department, McGill University, 3600 rue University, Montreal, QC H3A 2T8, Canada }

\vskip 0.3in

\end{center}

\vskip 0.35in

\begin{abstract} 
\noindent
The analytic structure of solutions to the Klein--Gordon equation in a black hole background, as represented by monodromy data, is intimately related to black hole thermodynamics.  It encodes the ``hidden conformal symmetry'' of a non-extremal black hole, and it explains why features of the  inner event horizon appear in scattering data such as greybody factors.  This 
indicates that hidden conformal symmetry is generic within a universality class of black holes. 
\end{abstract}

\vfill

\noindent \today

\end{titlepage}
}

\newpage

\section{Introduction}

The analytic structure of a black hole geometry encodes important information about its thermodynamic and quantum mechanical properties.  Most famously, geometric properties of the Euclidean section of a black hole metric encode its temperature and entropy \cite{Gibbons:1976ue}.  It is natural to ask whether other features of the analytic structure of black hole geometries encode important thermodynamic properties.  We will describe a relation between one such feature---the monodromy data of the Klein--Gordon operator---and the thermodynamic properties of the conjectured description of black hole microstates by a two-dimensional CFT.

Eternal black hole solutions contain numerous mathematical features, including past event horizons and multiple asymptotic boundaries, which are not present in realistic black holes  formed from collapse.  These mathematical features nevertheless appear to play an important role in the precise description of generic black hole microstates.  Recently a great deal of attention has focused on one such feature: the inner horizon of a black hole.  The inner horizon is typically unstable and  inaccessible to an external observer, yet plays an important role in the microscopic interpretation of the Bekenstein--Hawking entropy. Indeed, the fact that the inner horizon is a dynamical object with its own thermodynamic interpretation is a consequence of the standard description of black hole microstates in terms of a conformal field theory.  

To see this, we recall that the Bekenstein--Hawking entropy of any four- or five-dimensional, non-extremal, asymptotically flat black hole of Einstein--Maxwell theory can be written as
\bea\label{eq:BHE}
 S_+&=& {A_{+}\over 4G}= 2\pi \sqrt{{c\over 6} E_L} + 2\pi\sqrt{{c\over 6} E_R}~,
\eea
where $A_+$ is the area of the outer horizon, and $E_{L,R}$ and $c$ are defined appropriately \cite{Strominger:1997eq,Larsen:1997ge, Guica:2008mu, Castro:2010fd}. This coincides with the Cardy formula for the density of states of a two-dimensional CFT.
In the thermodynamic limit, where this formula applies, the left ($E_L$) and right ($E_R$) movers are decoupled.  This is a characteristic feature of the description of black hole microstates in terms of a CFT.
It follows that each of the individual terms on the right hand side of \eqref{eq:BHE} can be separately interpreted as a thermodynamic entropy.  Thus the linear combination
\bea
S_-= {A_{-}\over 4G}&=& 2\pi \sqrt{{c\over 6} E_L} -2\pi\sqrt{{c\over 6} E_R}~.
\eea
will obey the usual thermodynamic relations, including a first law.  This linear combination is precisely the area of the inner horizon, which has been observed to obey its own first law of black hole mechanics and a Smarr relation \cite{Cvetic:1997uw,Cvetic:1997xv,Castro:2012av,Chen:2012mh}.  The fact that the inner horizon, while physically unobservable, obeys standard thermodynamics relations is an indication
of the validity of the CFT description. For recent reviews, see \cite{Bredberg:2011hp,Compere:2012jk}.

Another key property is that the product of the areas $S_+ S_-$ is independent of the mass of the black hole \cite{Curir1,Curir2,Ansorg:2008bv,Ansorg:2009yi,Cvetic:2010mn,Castro:2012av}.  This universal property, while mysterious from a gravitational point of view, has a clear CFT interpretation: it reflects the fact that the left and right moving sectors have the same central charge \cite{Chen:2012mh,Chen:2013rb}.  It therefore provides further evidence for the CFT description.   At this point there is no completely satisfactory explanation of the role of the inner horizon, which in part can be tied to the lack of evidence for the CFT description.

The goal of the present work is to clarify the role of the inner horizon by revisiting the computation of greybody factors.  To do so, we will study the solutions of the scalar wave equation in a Kerr black hole background. We will pay special attention to the global analytic properties of the solutions, and to how this information is translated into physical observables. This technique will allow us to  address the question of why the split into left- and right-moving sectors is justified.  

We will see that analytic properties of Kerr wave functions are easily encoded in certain monodromy data of the Klein-Gordon equation. In the spirit of \cite{Motl:2002hd,Motl:2003cd,Neitzke:2003mz,Krasnov:2004ki}, we attribute a simple physical interpretation to these data by associating a particular basis of wave functions to each singular point of the Kerr wave equation. This will naturally explain the applicability of the ``hidden conformal symmetry'' argument to a wider class of black holes than simply Kerr.  In particular, we will argue that any black hole with three singular points will possess a similar hidden conformal symmetry. 

In the next section, we describe the monodromy data associated with the Kerr black hole and how to use it to give a simplified derivation of scattering coefficients.  The advantage of this approach is that we can derive a universal formula for certain types of scattering coefficients (related to greybody factors) without ever needing to explicitly solve the Kerr wave equation.  Several of the technical results quoted in this section 
will be explained in greater detail in \cite{us}.  In section \ref{sec:bhmono}, we will use this to provide a more systematic discussion of the hidden conformal symmetry structure of the Kerr black hole.  We will argue that there is a particular observer---defined in terms of a certain monodromy operator---whose low energy scattering cross sections match those of a CFT.  This provides insights into the applicability (and limitations) of the CFT description of non-extremal black holes.

\section{Black Hole Monodromy and Scattering Data}\label{sec:KerrMon}

We will now describe certain global analytic properties---monodromies---of solutions of wave equations in black hole backgrounds and their relations to physical observables.  
In this section, we will focus on the Kerr black hole probed by a massless scalar field $\psi$, though our results easily extend to other black holes and to fields with other mass and spin. 
Our discussion will be somewhat brief;  for a more complete discussion, see \cite{us} and references within.

\subsection{The Wave Equation}

We consider the Kerr metric in Boyer--Lindquist coordinates
\be\label{app:geom}
ds^2={\Sigma\over {\Delta}}dr^2- {{\Delta}\over \Sigma}\left(dt -{a}\sin^2\theta\, d\phi\right)^2+ \Sigma d\theta^2+{ \sin^2\theta \over \Sigma} \left((r^2+{a}^2)\,d\phi-{{a}}\,dt\right)^2~,
\ee
where  
\be
\Delta=r^2 +{a}^2-2M r~,\qquad \Sigma=r^2+{a}^2\cos^2\theta~.
\ee
The mass of the black hole is $M$ and the angular momentum is $J=Ma$.  We will write 
\be
\Delta=(r-r_+)(r-r_-)~, \qquad
{r}_\pm = M \pm \sqrt{M^2-{a}^2}~,
\ee
where $r_+$ and $r_-$ are the locations of the outer and inner horizons, respectively.

The metric has two isometries, allowing us to expand the field $\psi$ into modes
\be\label{bb:eigen}
\psi(t,r,\theta, \phi) = e^{-i\omega t+im\phi}R(r)S( \theta)~.
\ee
%We'll suppress the $\omega$ and $m$ arguments of $R$ and $S$ from now on.
Acting on these eigenmodes, the Klein--Gordon equation  separates \cite{Carter1968,Teukolsky:1972my} into a spheroidal equation\footnote{Further details on the spheroidal equation \eqref{bb:angeqn}, and in particular how to compute the eigenvalue $K_\ell$, can be found in \cite{us} and references within.}
\be\label{bb:angeqn}
\left[ {1\over \sin\theta}\partial_\theta
\left(\sin\theta\partial_\theta\right)-{m^2\over\sin^2\theta}+\omega^2 a^2\cos^2\theta \right] S(\theta)=-K_\ell  S(\theta)~,
\ee
and a radial equation
\be\label{bb:10}
\Big[\partial_r \Delta(r) \partial_r -V(r) \Big] R(r) = 0~,
\ee
where the effective potential  
\bea
 -V(r)={(\omega-\Omega_+ m )^2\over 4\kappa_+^2}{(r_+-r_-) \over (r-r_+)}   -{(\omega-\Omega_- m )^2\over 4\kappa_-^2}{ (r_+-r_-)\over (r-r_-)} +(r^2+2M(r+2M) )\omega^2-K_\ell~
\eea
is defined in terms of the surface gravity and angular velocity  at the horizons $r=r_\pm$:
\be
\kappa_{\pm}= {r_+-r_-\over 4M r_\pm}~,\qquad \Omega_\pm ={a\over 2Mr_{\pm}}~.
\ee

To understand the global properties of solutions, we consider \eqref{bb:10} as an equation in the complex $r$ plane (or more precisely, on the Riemann sphere $\mathbb{C}^*=\mathbb{C}\cup \{\infty\}$).
Equation \eqref{bb:10} has two regular singular points at $r=r_+$ and $r=r_-$.  Solutions to \eqref{bb:10} will have branch cuts at these points, as can be seen by constructing an appropriate series expansion around these points.  
For instance, around  $r=r_+$ we have two linearly independent solutions
\be
\label{n1kerr}
R_+^{out}(r)  = (r-r_+)^{i\alpha_+} \big( 1 + O(r-r_+)\big)~,\qquad
R_+^{in}(r)  = (r-r_+)^{-i\alpha_+} \big( 1 + O(r-r_+)\big)~,
\ee
where  
\be\label{bb:residue+}
\alpha_+ := {(\omega-\Omega_+ m )\over 2\kappa_+}~.
\ee
These solutions describe waves that are outgoing/ingoing at the horizon $r=r_+$, respectively, when $\omega\alpha_+>0$ (their roles are reversed when $\omega\alpha_+<0$).
From the wave equation we can compute subleading terms in \eqref{n1kerr} order by order in $(r-r_+)$; this provides a series expansion for $R(r)$ that converges when $|r-r_+| < |r_--r_+|$.
Near the the inner horizon ($r=r_-$) we have a similar expansion, where $\alpha_+$ is replaced by
\be\label{bb:residue-}
\alpha_- := {(\omega-\Omega_- m )\over 2\kappa_-}~.\ee

There is also a singular point at $r=\infty$,  near which we can construct a similar pair of series expansions 
\be
\label{n2kerr}
R^{out}_\infty(r) \sim e^{i \omega r} r^{ i \lambda-1} \big(1+ O(r^{-1})\big)~,\qquad
R^{in}_\infty(r) \sim e^{-i \omega r} r^{- i \lambda-1} \big(1+ O(r^{-1})\big)~,
\ee
 where 
\be
 \lambda =  2M \omega ~.
\ee
The subleading terms in \eqref{n2kerr} can be computed order by order in $r^{-1}$.
Unlike the singular points at $r=r_\pm$, however,  the singular point at $r=\infty$ is irregular. 
This means that the series \eqref{n2kerr} is asymptotic rather than convergent.  The series expansion for $R(r)$ appearing in \eqref{n2kerr} is therefore referred to as a formal solution to the wave equation, as opposed a true solution defined by an analytic continuation of the series expansion \eqref{n1kerr} to the whole complex $r$-plane.

\subsection{The Monodromy Data}

The analytic structure of the solutions can be understood more formally as follows.  
We first rewrite  \eqref{bb:10} as a pair of first order ODEs:
\be\label{bb:12}
\big(\partial_r -A_r(r) \big)  \left(\begin{array}{c} R(r)\\ \hat R(r)\end{array}\right) =0  ~,\qquad
A_r(r)= \left(\begin{array}{cc}0 & {\Delta(r)}^{-1} \\ V(r)& 0 \end{array}\right) ~,
\ee
which implies that
$\hat R= \Delta(r) \partial_rR(r)$. 
Formally, $A(r):=A_r(r) dr$ can be regarded as a flat $SL(2;\bb{C})$ connection on the complex plane with non-trivial holonomy around the singular points at $r=r_\pm, \infty$. 

We will denote by $(R_{1,2},\hat R_{1,2})^T$ a pair of linearly independent solutions to \eqref{bb:12}.  
 These can be combined into a matrix
\be
\Phi(r): =  \left(\begin{array}{cc} R_1(r)&R_2(r)\\ \hat R_1(r)&\hat R_2(r)\end{array}\right) ~
\ee
known as a fundamental matrix, which is also annihilated by the differential operator $\partial_r-A_r(r)$.  
The linear independence of $R_1$ and $R_2$ implies that $\Phi(r)$ is invertible. Of course, the choice of basis $R_{1,2}$ is not unique; one can go to a different basis of solutions by multiplying $\Phi(r)$ by some constant matrix from the right: $\Phi \to \Phi g$ for some $g \in GL(2,\C)$.

 Solutions $\Phi(r)$ can also be represented as path ordered exponentials
\be
\Phi(r) = \mathcal{P} \exp\left\{ \int_{r_0}^r A \right\} \Phi(r_0)~.
\ee
%Because the solutions to our wave equation have branch cuts, the resulting $\Phi(r)$ will depend on the homotopy class of the integration contour between $r_0$ and $r$.  
To understand the branching structure of $\Phi(r)$, consider transporting it around a closed loop $\gamma$ in the complex $r$-plane.  The result, which we will call $\Phi_\gamma(r)$, is also a fundamental matrix to our wave equation, so we must have that
\be
\label{eqn:monodef}
\Phi_\gamma(r) =  \mathcal{P}\exp\left\{ \oint_\gamma A \right\}  \Phi(r) =: \Phi(r) M_\gamma \, 
\ee
for some invertible constant matrix $M_\gamma$.  
This reflects the fact that a given a pair of solutions $R_{1,2}(r)$ will, after being analytically continued around this loop, return to themselves up to an overall linear transformation. 

The matrix $M_\gamma$ measures the lack of meromorphicity of $\Phi(r)$ and is called a \emph{monodromy matrix}.
If $\gamma$ does not enclose any branch points, then $M_\gamma=\mathbf{1}$.
 As $\gamma$ is deformed in the complex $r$-plane within a given homotopy class (i.e., without crossing any branch points), the monodromy matrix will not transform (though, ${\cal P}e^{\oint_\gamma A}$ will transform by conjugation if the base point is moved).  Thus, the monodromy matrices form a representation of the fundamental group of the complex $r$-plane with punctures at the singularities of the wave equation.

It is easy to determine the Jordan normal form of certain monodromy matrices.
Near $r=r_+$, \eqref{n1kerr} implies that our fundamental matrix can be expanded as
\be\label{aa:regexpanP}
\Phi(r) = \bigg( \sum_{n=0}^{\infty} (r-r_+)^n \Phi_n^+ \bigg)
\left(\begin{array}{cc} (r-r_+)^{i\alpha_+} & 0 \\ 0 & (r-r_+)^{-i\alpha_+} \end{array} \right) g_+ \ 
\ee
for some constant matrix $g_{+}\in GL(2,\C)$, and where the $\Phi_n^+$ are constant matrices with $\Phi_0^+\in SL(2,\C)/\C^*$.\footnote{$\C^*$ is generated by $e^{\eta \sigma^3}$, where $\eta\in\C$ and $\sigma^3$ is a Pauli matrix.  We could let $\Phi_0^+$ take values in $GL(2,\C)$, but some of these $GL(2,\C)$ matrices could be absorbed into the definition of $g_+$, making them redundant.}  This determines the monodromy matrix $M_+$ for a loop circling only the singularity at $r_+$ exactly once.  So $M_+$ is
\be\label{eq:matrices1}
M_{+}  = g_+^{-1} \left(\begin{array}{cc}e^{-2\pi  \alpha_+} &0\\ 0&e^{2\pi  \alpha_+} \end{array}\right) g_+  \cong \left(\begin{array}{cc}e^{-2\pi  \alpha_+} &0\\ 0&e^{2\pi  \alpha_+} \end{array}\right) \, ,
\ee
where `$\cong$' denotes that both sides are conjugate to each other as two-by-two matrices.  Similarly, the monodromy matrix $M_-$ of a loop circling only $r_-$ exactly once obeys
\be
 M_{-}  \cong \left(\begin{array}{cc}e^{-2\pi  \alpha_-} &0\\ 0&e^{2\pi  \alpha_-} \end{array}\right)\, ,
\ee
except in the case of Schwarzschild where $\alpha_-=0$ and
\be\label{eq:ressch}
M_- \cong \left(\begin{array}{cc} 1 & 1 \\ 0 & 1 \end{array}\right) ~ .
\ee

Computing the monodromy matrix $M_{\infty}$ around the irregular singular point at $r=\infty$ is somewhat more subtle.  In general, of course, it will take the form\footnote{{Assuming that $M_\infty$ is diagonalizable.  Similar to $M_-$ in \eqref{eq:ressch}, when $e^{4\pi\alpha_\irr}=1$ there can be a nontrivial Jordan block.}}
\be\label{eq:matrices2}
M_{\infty}  \cong \left(\begin{array}{cc}e^{-2\pi  \alpha_\irr} &0\\ 0&e^{2\pi  \alpha_\irr} \end{array}\right) \, ,
\ee
for some parameter $\alpha_\irr$.  If the series \eqref{n2kerr} were convergent, then we would simply identify $\a_\irr$ with $\lambda$.  
Unfortunately, since \eqref{n2kerr} is only an asymptotic series it is not as simple as that to compute $\a_\irr$.
Instead, as a consequence of Stokes phenomenon, the true solutions to the wave equation will have a more complicated branch structure than appears in the asymptotic series expansion \eqref{n2kerr}.  A full discussion of this, including a calculation of $\alpha_\irr$, will appear in \cite{us}.  For now, however, we will simply use $\alpha_\irr$ to parameterize the non-trivial monodromy around $r=\infty$.

Finally, we note that a loop that simultaneously encloses all three singularities is trivial.  Thus
\be\label{dd:0}
M_-M_+M_\infty=\mathbf{1}~.
\ee 
This relation will explain why inner horizon data appears in black hole scattering and CFT computations.

\subsection{Scattering Data and Greybody Factors}
\label{sec:scattering}

We now explain the relationship between these monodromy matrices and scattering coefficients.  The essential observation is that any scattering
computation involves a change of basis between solutions with boundary conditions fixed in one region of the geometry and those that have boundary conditions fixed in another region.  Each region will be associated with a singular point of the wave equation, so this change of basis can be characterized in terms of the left eigenvectors of the relevant monodromy matrices.

Let us illustrate this for the Kerr black hole, where we are interested in studying the scattering of a scalar wave sent from asymptotic infinity off of the black hole horizon.   
Consider a basis of solutions $R_{1,2}$, with associated fundamental matrix $\Phi$.  We will denote by $M_\pm$ and $M_\infty$ the monodromies around $r=r_\pm, \infty$ in this basis.
Near the outer horizon, $R_{1,2}$ will be linear combinations of the ingoing and outgoing  wave functions $R_+^{in}$ and $R_+^{out}$.  In terms of fundamental matrices, 
 \be\label{eq:PG+}
 \Phi = \Phi_+ g_{+} ~ ,
 \ee 
 where $\Phi_+$ is the fundamental matrix built out of $R_+^{in/out}$ and $g_+\in GL(2,\mathbb{C})$.
In the $R_+^{in/out}$ basis, the monodromy $M_+$ is diagonal.  Thus $g_+$ is a matrix that diagonalizes $M_+$, i.e., its rows are left eigenvectors of $M_+$.  
Similarly, we can consider a basis of solutions $\Phi_\infty$ defined by 
\be
\Phi = \Phi_\infty g_\infty ~ ,
\ee
where $g_\infty\in GL(2;\C)$ contains the left eigenvectors of $M_\infty$ ($\Phi_\infty$ is known as a Floquet solution).\footnote{In case $M_\infty$ is not diagonalizable, we can take $g_\infty$ to contain the \emph{generalized} eigenvectors of $M_\infty$.}

To compute generalized scattering data, we must find the change of basis between the solutions $\Phi_\infty$ and those that are ingoing/outgoing at $r=r_+$, $\Phi_+$.  That is to say, we want to compute 
\be\label{eq:conn-def}
{\cal M}_{\infty\to +}=\Phi_\infty^{-1} \Phi_+ = g_\infty g_+^{-1}~.
\ee
The matrix ${\cal M}_{\infty \to +}$, called the connection matrix, contains the generalized scattering data.

To relate this to more traditional treatments of scattering problems, we need to fix a normalization for our various bases of solutions.  At the outer horizon, our choice is equivalent to the standard plane wave boundary conditions in tortoise coordinates.  If at infinity we also were to use a basis of plane waves (i.e., the canonical definition of ingoing/outgoing), instead of what we will use, we would  normalize our solutions using the standard signature $(1,1)$ inner product induced by the Klein--Gordon norm.  This would require ${\cal M}_{\infty \to +}$ to be an $SU(1,1)$ matrix, so it would take the familiar form  
\be\label{aa:99}
\left(\begin{array}{cc} \frac{1}{{\cal T}}  &  \frac{{\cal R}}{{\cal T}} \\ \frac{{\cal R}^*}{{\cal T}^*} & \frac{1}{{\cal T}^*} \end{array}\right)~,\qquad |{\cal R}|^2+|{\cal T}|^2=1~,
\ee
for some complex constants ${\cal T}$ and ${\cal R}$.  These would then be the reflection and transmission coefficients for the canonical scattering problem.  The norm $|{\cal T}|^2$ would be what is known as the greybody factor, representing the correction to the pure blackbody spectrum of Hawking radiation due to scattering off the gravitational potential.

Whether it is possible to make ${\cal M}_{\infty \to +}$ \eqref{eq:conn-def} an $SU(1,1)$ matrix  depends on the monodromies $e^{2\pi\alpha_i}$.  In general, for a scattering problem along  the real $r$ axis, it is sufficient  to demand that  all $e^{2\pi\alpha_i}$ are real and no entires of ${\cal M}_{\infty\to +}$ vanish---this will be evident from the explicit expression below. 
For real values of the parameters in an ODE,  the monodromies $e^{2\pi\alpha_i}$ will be either real or phases,\footnote{For real ODE, if $R(r)$ is a solution then so is $R^*(r)$.  If  $R(r)$ has monodromy $e^{2\pi\alpha_i}$, then $R^*(r)$ has monodromy $e^{2\pi\alpha_i^*}$.  Since the two eigenvalues of $M_i$ are $e^{\pm2\pi\alpha_i}$ and $R^*(r)$ is a solution of the ODE, $e^{2\pi\alpha_i}$ must be real or a phase.} and for physical ranges of the parameters in our ODE \eqref{bb:10} ($\omega\in \R$, $r_+>r_-$, etc.), we see that $\alpha_\pm \in \R$.

However, it turns out that $e^{2\pi\alpha_\irr}$ can be real or a phase, depending on the frequency $\omega$ \cite{us}.  When $e^{2\pi\alpha_\irr}$ is real, ${\cal M}_{\infty\to +}$ will have the form \eqref{aa:99}.  When it is a phase, ${\cal M}_{\infty\to +}$ will instead take the two-by-two matrix of Klein--Gordon inner products of the two solutions in $\Phi_\infty$, which is $\sigma^2$, to the matrix of inner products of the two solutions in $\Phi_+$, which is $\sigma^3$ ($\sigma^{2,3}$ are Pauli matrices).  To encompass both cases, we simply write
\be
\label{eq:Mform}
{\cal M}_{\infty \to +}= 
\left(\begin{array}{cc} \frac{1}{{\cal T}}  &  \frac{{\cal R}}{{\cal T}} \\ \frac{{\cal R}'}{{\cal T}'} & \frac{1}{{\cal T}'} \end{array}\right)~,\qquad {\cal R}{\cal R}'+{\cal T}{\cal T}'=1~,
\ee
where $e^{2\pi\alpha_\irr}$ real corresponds to the identifications ${\cal R}' ={\cal R}^*$ and ${\cal T}' = {\cal T}^*$, while $e^{2\pi\alpha_\irr}$ a phase corresponds to the identifications ${\cal R} = i\frac{{\cal T}}{{\cal T}^*}$ and ${\cal R}' = i\frac{{\cal T}'}{{\cal T}'^*}$.

To compute ${\cal M}_{\infty\to+}$ we need to known not just the conjugacy classes of $M_+$ and $M_\infty$, but their actual entries in a common basis.  For the Kerr black hole, since there are only three singular points, this is a simple problem in linear algebra.  Explicitly, given the conjugacy classes of the monodromy matrices  
\be
\det (M_i) =1\, , \qquad {\rm tr}( M_i) = 2\cosh(2\pi  \alpha_i)\, , \qquad  M_i \neq \mathbf{1} \, ,  \qquad  \textrm{for}~~~ i=-,+,\infty\, ,
\ee
and equation \eqref{dd:0},
\be
M_- M_+ M_\infty \ = \ \mathbf{1}~,
\ee
the monodromy matrices can be determined.  Up to an overall conjugation, they are
\begin{gather}\label{eq:basisM}
 M_-=\left(\begin{array}{cc}0&-1\\ 1&2\cosh(2\pi  \alpha_-)\end{array}\right) ~,\qquad\qquad M_+= \left(\begin{array}{cc}2\cosh(2\pi  \alpha_+) &e^{2\pi  \alpha_\irr}\\ -e^{-2\pi  \alpha_\irr}&0\end{array}\right)~,  \\
 \label{eq:basisM2}
M_\infty=\left(\begin{array}{cc}e^{2\pi  \alpha_\irr}&0\\ 2\big(e^{-2\pi  \alpha_\irr}\cosh(2\pi  \alpha_-)-\cosh(2\pi  \alpha_+) \big)&e^{-2\pi  \alpha_\irr}\end{array}\right)~.
\end{gather}
The connection matrix can be constructed from the eigenvectors of $M_\infty$ and  $M_+$.  It is
\be
\label{eq:scattering}
{\cal M}_{\infty\to +} \ = \  \left(\begin{smallmatrix} d_1 &  \\  & d_2 \end{smallmatrix}\right)  \left(\begin{array}{cc} \sinh\pi(\alpha_\irr-\alpha_++\alpha_-) & \sinh\pi(\alpha_\irr+\alpha_++\alpha_-)  \\  \sinh\pi(\alpha_\irr+\alpha_+-\alpha_-)  &  \sinh\pi(\alpha_\irr-\alpha_+-\alpha_-)  \end{array}\right)  \left(\begin{smallmatrix} d_3 &  \\  & d_4 \end{smallmatrix}\right)  \, ,
\ee
for some unknown constants $d_i$.  These constants appear because, in constructing a change of basis matrix, we are always free to rescale
our eigenvectors by arbitrary constants.

 Bringing ${\cal M}_{\infty \to +}$ to the form \eqref{eq:Mform} fixes some combinations of the unknown constants $d_i$. Even without knowing the $d_i$, by taking appropriate ratios we obtain
\be
\label{eq:simple-transmission}
{\cal T}{\cal T}' = 1-{\cal R}{\cal R}' =  \frac{\sinh(2\pi\alpha_+)\,\sinh(2\pi\alpha_\irr)}{\sinh\pi(\alpha_-+\alpha_+-\alpha_\irr)\, \sinh\pi(\alpha_--\alpha_++\alpha_\irr)} \, ,
\ee
which is the relative flux between the horizon and the basis at infinity.  There is still some ambiguity in the $d_i$ that cannot be fixed by this simple procedure,\footnote{The ambiguity is actually important for the computation of quasinormal modes. See \cite{us} for further discussion on this topic.} but for our purposes here this is sufficient.

This expression for ${\cal T}{\cal T}'$ contains important information: it highlights the dependence of scattering coefficients on analytic properties of the black hole geometry. In particular, it explains why inner horizon data appears in scattering computations.  For scattering processes between  an outer horizon and infinity, one would naively expect that the answer cannot depend on properties of the inner horizon.   However,   
equation \eqref{dd:0} clearly implies that outside the horizon scattering process can probe internal features of the black hole geometry, such as the monodromy $\alpha_-$.
This might appear problematic, as reflection and transmission coefficients are directly accessible to observers at asymptotic infinity and thus should not contain information about inside the horizon physics.  However, we must remember that we are dealing with eternal black holes, which contain many features (such as white hole singularities and past event horizons)
that will not appear in physical black holes formed from collapse. Such black holes, of course, would not have such a simple analytic structure.

This discussion might give the impression that we have solved exactly the scattering problem for the eternal Kerr black hole, but this is not the case.  
The basis $\Phi_\infty$ considered above is associated with left eigenvectors of the monodromy matrix $M_\infty$, which are not asymptotic to purely ingoing or outgoing plane waves at infinity, as described in \eqref{n2kerr}, but to linear combinations of them. The scattering coefficients relevant for astrophysics would come from the connection matrix between the plane wave basis and $\Phi_+$.  This more complicated problem will be considered in \cite{us}.  Our result \eqref{eq:simple-transmission} should be viewed as a variant of the canonical scattering computation.

We note  that for black holes in asymptotically AdS spaces, the singularity at infinity is regular and this subtlety is not present.  Thus, for example, this procedure gives a completely correct computation of the scattering coefficients of the BTZ black hole.  Indeed, many attempts to understand microscopic properties of black holes involve discarding certain features of solutions near asymptotic infinity.  In particular, one typically takes some sort of near horizon limit that replaces flat asymptotic region with that of AdS or a similar geometry.  Our present approach, where we construct scattering states from $\Phi_\infty$, can be viewed as a particular implementation of this general strategy.

\section{Black Hole Monodromy and CFT}
\label{sec:bhmono}

We now turn to the thermodynamic interpretation of black hole monodromy.  We will see that the transmission coefficient \eqref{eq:simple-transmission} can be interpreted as the finite temperature absorption cross section of the two-dimensional CFT, and that the monodromy matrices can be regarded as defining sets of wave functions associated with this CFT.

\subsection{Monodromies and Local Observers}

We will first describe the basic relationship between monodromy and gravitational thermodynamics.  Consider a loop $\gamma$ in the complex $r$ plane.  Assuming that the monodromy matrix $M_\gamma$ is diagonalizable, we can seek a basis of solutions that do not mix when transported around the loop $\gamma$.     In terms of our original basis $R_{1,2}$, these solutions will take the form $v_1 R_1 + v_2 R_2$, where $(v_1,v_2)$ is a left eigenvector of  $M_\gamma$.  This new basis of solutions can be used to define a vacuum state in the quantum field theory of the field $\psi$ in the fixed black hole background.

In order to interpret this basis of solutions, let us consider the simple example of two-dimensional flat space in Rindler coordinates  
\be
ds^2 = dr^2 - r^2 d\tau^2~,
\ee
Expanding in a basis of modes $e^{-i\omega_\tau\tau}\psi_{\omega_\tau}(r)$ with fixed frequency $\omega_\tau =-i \p_\tau$, the scalar wave equation implies that modes satisfy
\be
\left({1\over r}\p_r(r \p_r) + {\omega_\tau^2 \over r^2}\right) \psi_{\omega_\tau}(r) = 0 ~ , 
\ee
which has regular singular points at $r=0,\infty$.  Of course, the simple pole at $r=0$ is precisely the Rindler horizon where $|\partial_\tau|$ vanishes.  

The eigenfunctions will exhibit monodromy transformations as they are analytically continued in the complex $r$ plane around $r=0$.   The  ingoing/outgoing solutions at the Rindler horizon $r=0$ have the form $r^{\pm i \omega_\tau}$, and so 
pick up phases $e^{\mp 2\pi \omega_\tau} $ as we take $r\to e^{2\pi i} r$.  Thus the monodromy matrix $M_\gamma$ for the infinitesimal loop around $r=0$ will have eigenvalues $e^{\pm 2 \pi \omega_\tau}$.  The eigenvectors of $M_\gamma$ will determine the ingoing/outgoing wave functions at $r=0$.

Further, note that on the eigenmode $e^{-i\omega_\tau \tau}\psi_{\omega_\tau}(r)$, the monodromy transformation acts trivially if time $\tau$ is identified with $\tau+2\pi i$, which is precisely the periodicity in imaginary time that we expect for the thermal Green's function in a Rindler background.  So we observe that the thermal Green's function is actually invariant under monodromy transformations around the horizon.  Hence the monodromies can be used to easily determine the thermal properties of the Green's function.

To summarize, we see that when the curve $\gamma$ circles a Killing horizon the matrix $M_\gamma$ has a clear physical interpretation:
\begin{itemize}
\item
The eigenvectors of $M_\gamma$ are the vacuum wave functions of Rindler observers at the horizon in a given basis;
\item
The eigenvalues of $M_\gamma$ are the exponentials of the local Rindler energies $e^{\pm2 \pi \omega_\tau}$; and,
\item 
Invariance of the thermal Green's function under imaginary coordinate identifications is equivalent to invariance under monodromy transformations from circling the dual Killing horizon in Rindler-like coordinates.
\end{itemize}

\subsection{Hidden Conformal Symmetry}\label{sec:HCS}

We will now apply this logic to the Kerr solution, which has a pair of Killing horizons.  We will consider the basis of wave functions that are eigenstates of $M_\infty$, the monodromy around $r=\infty$.  These can alternately be viewed as eigenstates of the product $(M_-M_+)^{-1}$.  The monodromies $\alpha_\pm$ of the inner and outer horizon in \eqref{bb:residue+} and \eqref{bb:residue-} should be interpreted as energies associated with a particular set of states.  For reasons that will soon become apparent, we consider the following linear combinations of the energies associated with the two Killing horizons:
\bea\label{eq:omegaLR}
 \omega_L &:=& \alpha_+ - \alpha_-~,\cr
 \omega_R&:=& \alpha_++ \alpha_-~,
\eea
These energies are functions of the eigenvalues $(\omega, m)$ of the operators $(i\partial_t, -i\partial_\phi)$.
The energies $\omega_{L,R}$ are conjugate to the two variables $t_{L,R}$, with $(\omega_{L},\omega_R)$ eigenvalues of $\left(i\partial_{t_{L}},i\partial_{t_R}\right)$.  Thus, 
\be
e^{-i\omega t +i m \phi}=e^{-i{\omega_L} t_L -i{\omega_R} t_R }~.
\ee
Using the explicit form of the monodromies for Kerr \eqref{bb:residue+}--\eqref{bb:residue-}, we find that 
\bea\label{monbasis}
t_R =2\pi T_R \,\phi~,\qquad 
t_L = {1\over 2M}t - 2\pi T_L\, \phi~,
\eea
where $(t,\phi)$ are Boyer--Lindquist coordinates, and 
\be
T_L= {r_++r_-\over 4\pi a}~,\qquad T_R= {r_+-r_-\over 4\pi a}~,
\ee
which can be rewritten in a way convenient for  generalization to other solutions as
\be\label{eq:TKO}
 T_L= {1\over 2\pi}{\kappa_- +\kappa_+\over \Omega_--\Omega_+}~,\qquad T_R={1\over 2\pi} {\kappa_-- \kappa_+\over \Omega_--\Omega_+}~.
 \ee
Note that $\p_{t_L}+\p_{t_R}$ and $\p_{t_L}-\p_{t_R}$ are the Killing vectors that vanish on the outer and inner horizons, respectively. 

The  identification  of the angular coordinate $\phi\sim\phi +2\pi$ leads to 
\be\label{eq:TC}
(t_L, t_R) \sim (t_L,t_R)+4\pi^2 \big(-T_L,T_R)~.
\ee
This restricts the domain of the monodromy coordinates $t_{R,L}$ to an infinite strip.  We also note that the usual thermal identification
\be
(t,\phi) \sim (t,\phi) + \tfrac{2\pi i}{\kappa_+}\, (1,\Omega_+) ~
\ee
is equivalent to
\be\label{eq:TH}
(t_L, t_R)\sim (t_L,t_R)+2\pi i (1,1)~.
\ee
This identifies the eigenmodes $e^{-i\omega t+i m \phi}$ with multiplication by $e^{4\pi\alpha_+}$, which is the monodromy one obtains circling the outer horizon twice in the $r$-plane.  This is expected since the Rindler-like radial coordinate lives on a double-cover of the $r$-plane---i.e., near the horizon the Rindler-like radial coordinate is $u$, where $r \sim r_+ + u^2$.  We note that a similar identification with respect to the inner horizon, $(t,\phi) \sim (t,\phi) + \frac{2\pi i}{\kappa_-} (1,\Omega_-)$, would lead to an invariance of the Green's function under monodromy transformations from circling the inner horizon.

With these definitions, the transmission coefficient \eqref{eq:simple-transmission} becomes
\bea\label{eq:cT}
{\cal T}{\cal T}'
&=&\frac{\sinh2\pi( \omega_L +\omega_R)\,\sinh(2\pi\alpha_\irr)}{\sinh\pi(\omega_L-\alpha_\irr)\, \sinh\pi(\omega_R+\alpha_\irr)} ~.
\eea
Let us first consider this expression assuming that $\alpha_\irr$ has no dependence on $(\omega,m)$. 
Then \eqref{eq:cT} is the scattering coefficient of a two-dimensional CFT; it describes a thermal system with two decoupled left- and right-moving sectors with energies $\omega_{L,R}$ and temperatures $T_{L,R}$.
The monodromy coefficient $\alpha_\irr$ determines the conformal weight of the CFT operator dual to the scalar mode under consideration. 

 It is useful to compare this to the original hidden conformal symmetry 
proposal \cite{Castro:2010fd}. The approach there relied on a low frequency approximation and a ``near'' limit of the phase space variables, which was distinct from a geometrical near-horizon limit.  In this approximation, the temperatures $T_{L,R}$ were extracted and scattering coefficients were found to match with those of a CFT.  The present approach provides a derivation of these temperatures based on monodromy data, without the need for a low energy limit.  (We will compare to the related proposals of \cite{Chen:2012mh,Chen:2013rb}, which exploit the thermodynamics properties of the solutions, in section \ref{sec:mt}.)

In general, however, $\alpha_\irr$ has a complicated dependence on $\omega$ and $m$ that vanishes only in a low frequency limit.   As will be shown in \cite{us}, 
\be
 i\alpha_\irr = \ell - 2 M^2 \omega^2 \, \tfrac{15\ell(\ell+1)-11}{(2\ell+3)(2\ell+1)(2\ell-1)} +{O(\omega^3)}~, \qquad \textrm{(for $\ell\neq 0$)}
\ee
 with $\ell\in \Z^+$. Thus when $\omega M\ll 1$, $\alpha_\irr=-i\ell$ is a constant and we are in the same low frequency limit as \cite{Castro:2010fd}. 
 At finite $\omega$, however,  the black hole is not well approximated by a CFT.\footnote{Note that when $\alpha_\irr \in \frac{i}{2}\mathbb{Z}$, $M_\infty$ may contain a non-trivial Jordan block, in which case it would only have one eigenvector.  In this case, equations \eqref{eq:basisM}--\eqref{eq:basisM2} are still valid, but \eqref{eq:scattering} (and, hence, \eqref{eq:cT}) requires modification.  One should replace ${\cal M}_{\infty\to +}$ by the change of basis between eigenvectors of $M_+$ and \emph{generalized} eigenvectors of $M_\infty$.} 
 For example, the left- and right-moving sectors will not decouple at finite $\omega$.

We conclude that the hidden conformal symmetry is associated with a basis of solutions that are eigenstates of the monodromy operator $M_\infty = (M_- M_+)^{-1}$.  The ``observer'' whose scattering experiments agree with those of a CFT in the low energy limit is precisely the one whose vacuum wave functions do not mix under $M_\infty$.  These vacuum wave functions are precisely those appearing in $\Phi_\infty$ of section \ref{sec:scattering}.  It is important to note that this CFT observer is not the standard observer of asymptotically flat space.  The standard asymptotically flat observer defines vacuum wave functions that are plane-wave scattering states at $r=\infty$, i.e., those that are asymptotic to the formal solutions \eqref{n2kerr}.  From the point of view of a standard asymptotic observer, the wave functions of the CFT observer encode properties of the black hole in a highly non-trivial way associated with Stokes phenomena.

%%%%%%%%%%%%%%%%%%%%%%%%%%%%%%%%%%%%%%%%%%%%%%%%%%%%%%%%%%%%%%%
%%%%%%%%%%%%%%%%%%%%%%%%%%%%%%%%%%%%%%%%%%%%%%%%%%%%%%%%%%%%%%%%

\section{Discussion}\label{sec:mt}

We have described a relationship between monodromy data and black hole thermodynamics, which relies on the observation that to each singular point we can associate a particular set of states.  Our expression for the transmission coefficient \eqref{eq:cT} holds for any black hole with three singular points.  In particular, it applies regardless of the nature of the singular points---regular or irregular---so it can be used for black holes with different asymptotic boundary conditions. The transmission coefficient will naturally describe a pair of decoupled sectors associated with the two other singular points.  In particular, for any black hole with 3 singular points we will obtain a decomposition of the form \eqref{eq:cT} involving data symmetrically from the inner and outer horizons. We illustrate this explicitly in appendix \ref{app:B} for the five-dimensional Myers--Perry black hole.  Similar considerations also apply to the five-dimensional black ring---this is true despite the fact that in this case the wave equation is no longer separable.
  
 Indeed, the monodromy eigenvalues 
\be\label{dd:MM}
\alpha_\pm= {(\omega-\Omega_\pm m )\over 2\kappa_\pm}~,
\ee 
 are conjugate to the null Killing vectors that define the outer/inner horizon. 
 Thus, in general, we expect that $any$ regular singular point that arises due to a Killing horizon with non-vanishing surface gravity $\kappa$ will lead to a  monodromy of the form \eqref{dd:MM}.
A cosmological horizon falls into this category; we illustrate this explicitly in appendix \ref{app:kads} for Kerr--AdS.  The only difference is that the relevant poles no longer lie on the real $r$ axis.\footnote{The fact that cosmological horizons are on equal footing with Killing horizons was observed in \cite{Cvetic:2010mn,Birkandan:2011fr}.}  In contrast, we note that an extremal horizon leads to an irregular singular point.\footnote{One can think of irregular singularities as arising through the process of confluence, where multiple regular singular points coincide.  This happens in the extremal limit of a Kerr black hole, where $r_+\to r_-$.  It also happens for Kerr--AdS, where as $\Lambda\to 0$  the cosmological horizons combine to form the irregular singular point at infinity of flat space.}

It is worth pointing out the monodromies $\alpha_\pm$ have a second important physical interpretation. 
For a general bifurcate Killing horizon, Wald has noted that the entropy is equal to a Noether charge associated to a Killing field \cite{Wald:1993nt}.  For the Kerr black hole, the entropies $S_{\pm}$ are the charges associated with the Killing fields
\be
\zeta^{\pm}=\kappa_{\pm}( \partial_t +\Omega_\pm \partial_\phi) ~.
\ee
Let us imagine  probing such a black hole by a scalar field excitation with energy $\omega$ and angular momentum $m$.  This induces  infinitesimal changes $\delta M=\omega$ and $\delta J=m$ in the mass and angular momentum of the black hole.  
The variations in the entropies are then precisely the monodromies:
\be\label{dd:monS}
\delta S_{\pm} = \pm 4\pi\alpha_\pm~.
\ee
 This holds for any gravitational theory to which Wald's formula applies, including those with higher derivative terms and gravitational anomalies.  However, these infinitesimal relations do not imply global identities, such as Smarr relations. 
 
 Writing the monodromy in this manner makes evident that our definitions here are in complete agreement with  thermodynamic method developed in \cite{Chen:2012mh,Chen:2013rb}.
 To see this, consider again \eqref{dd:MM}. Solving for the orbital angular momentum we find 
 \bea\label{ff:aa}
 \delta J&=& {2\over \Omega_- -\Omega_+}(\kappa_+\alpha_+ -\kappa_-\alpha_-)
 = 2\pi T_L \omega_L - 2\pi T_R\omega_R
 \eea 
where we have used \eqref{eq:omegaLR} and \eqref{eq:TKO}.  We also have the phenomenological observation that the product
 \be\label{eq:ss}
 {S_+S_-\over 4\pi^2} = {\cal F}(J)~,
 \ee
is independent of the mass.  Varying \eqref{eq:ss} gives
  \bea\label{eq:ss2}
2\pi^2  {\partial {\cal F}\over \partial J} \delta J  &=& S_L \,\delta S_L - S_R \,\delta S_R=  2\pi S_L\, \omega_L - 2\pi S_R \, \omega_R~,
 \eea
  where we have introduced the notation
  \be
  S_\pm = S_L \pm S_R ~,\quad \delta S_{L,R}=2\pi \omega_{L,R}~.
  \ee
Comparing \eqref{ff:aa} with \eqref{eq:ss2} we find
\be\label{eq:CC}
S_{L,R}=  2\pi^2  {\partial {\cal F}\over \partial J} T_{L,R}~.
\ee
 It is  tempting to interpret this as the Cardy formula for the density of states of a two-dimensional CFT:
 \be
 S_{L,R}=  {\pi^2\over 3} c \, T_{L,R}~,
 \ee
with central charge% 
 \be
 c=  6 {\partial {\cal F}\over \partial J} ~.
 \ee
 The inner/outer horizon entropies can then be written in microcanonical ensemble as
 \be\label{ff:ee}
 S_{\pm}= 2\pi \sqrt{{c\over 6} E_L} \pm 2\pi\sqrt{{c\over 6} E_R}~,
 \ee
with 
\be
E_{L,R}= {\pi^2\over 6}c\, (T_{L,R})^2~.
\ee
It is important to emphasize that the ensemble described above is not the standard one associated with a conformal field theory with  decoupled left- and right-moving sectors.  In a CFT, one typically fixes the central charge and considers canonical ensembles where $T_{L,R}$ are varied independently.  The black hole, however, is associated with an ensemble where $T_L^2 - T_R^2 = {1\over 4\pi^2}$ is fixed.  This may indicate that the left- and right-moving sectors are coupled in some way.  This may be related to the fact that $\alpha_\irr$ depends on $(\omega, m)$ in a non-trivial way outside of the low energy limit.

%%%%%%%%%%%%%%%%%%%%%%%%%%%%%%%%%%%%%%%%%%%%%%%%%%%%%%%%%%%%%%%
%%%%%%%%%%%%%%%%%%%%%%%%%%%%%%%%%%%%%%%%%%%%%%%%%%%%%%%%%%%%%%%%

\section*{Acknowledgements}

{We thank Tatsuo Azeyanagi, St\'ephane Detournay, Gary Gibbons, Finn Larsen, Don Marolf,  and Andy Strominger, for useful conversations.  The work of A.M. and A.C. was in part supported by NSF under Grant No. PHY11-25915.  A.C.'s work is also supported by the Fundamental Laws Initiative of the Center for the Fundamental Laws of Nature, Harvard University. A.M. and J.M.L. are supported by the National Science and Engineering Research Council of Canada.  M.J.R. is supported by the European Commission - Marie Curie grant PIOF-GA 2010-275082.}

%%%%%%%%%%%%%%%%%%%%%%%%%%%%%%%%%%%%%%%%%%%%%%%%%%%%%%%%%%%
%%%%%%%%%%%%%%%%%%%%%%%%%%%%%%%%%%%%%%%%%%%%%%%%%%%%%%%%%%%%
\appendix

\section{Myers--Perry black hole}\label{app:B}

In this appendix, we extend our discussion to five-dimensional Myers--Perry black holes \cite{Myers:1986un}. We will see that monodromy data contains the information about left- and right-moving temperatures associated with this black hole's hidden conformal symmetry \cite{Castro:2010fd, Krishnan:2010pv}.  Our definitions and conventions follow  \cite{Cvetic:1997uw,Lu:2008jk,Krishnan:2010pv}.

The radial part of the wave equation for a massless scalar field is 
\bea
{\partial\over\partial x}\big(x^2-\tfrac{1}{4}\big){\partial\over\partial x}\Phi
&\!\!\!\!+\!\!\!\!&{1\over 4}\Bigg[x\Delta\omega^2-\Lambda+M\omega^2
\\
&\!\!\!\!& \quad +{1\over x-{1\over 2}}
{\left({\omega}-m_R {\Omega^R_+}
-m_L {\Omega^L_+}\right)^2\over \kappa_+^2}
-{1\over x+{1\over 2}}{\left({\omega}-m_R {\Omega_-^R}
-m_L {\Omega_-^L}\right)^2\over \kappa_-^2}
\Bigg]\Phi = 0~, \nonumber
\eea
where we have decomposed the scalar field as
\bea
\psi&=& R(x)\,\chi(\theta)\, e^{-i\omega t+im_R(\phi+\psi)+im_L(\phi-\psi)}~,
%
%R(x) ~\chi(\theta)~e^{-i\omega t+im_\phi\phi+im_\psi\psi}=
%\sum_{m_L,m_R} \int d\omega \, R(\omega,m_L,m_R;x)\,\chi(\omega,m_L,m_R;\theta)\, e^{-i\omega t+im_R(\phi+\psi)+im_L(\phi-\psi)}~,
\eea
%where from now on we drop the arguments $(\omega,m_L,m_R)$ from $R$ and $\chi$.  
See \cite{Cvetic:1997uw} for the definition of the eigenvalue $\Lambda$. The variable $x$ is related to the Boyer--Lindquist coordinates by
\bea
x \equiv {r^2 - {1\over 2}(r^2_{+}+r^2_{-})\over
(r^2_{+}-r^2_{-})}~.
\eea
Here $r_\pm$ are the locations of the inner and outer horizons, which are determined by the zeroes of
\be
\Delta = (r^2+a^2)(r^2+b^2)-2Mr^2 = (r^2-r_+^2)(r^2-r_-^2)~.
\ee

The angular velocities and surface gravities are
\bea
\Omega^L_{\pm}&=&{r_+^2\over  r_\pm}\frac{(a-b)(r_+-r_-)}{(r_+^2+a^2)(r_+^2+b^2)}~,  \cr  \Omega^R_{\pm}&=&{r_+^2\over r_\pm}\frac{(a+b)(r_++r_-)}{(r_+^2+a^2)(r_+^2+b^2)} ~,\cr
\kappa_\pm&=&{r_+^2\over r_\pm}\frac{(r_+^2-r_-^2)}{(r_+^2+a^2)(r_+^2+b^2)}~,
\eea
and 
\be
\Omega_\pm ^\phi ={1\over 2}\left(\Omega^R_{\pm}\pm \Omega^L_{\pm}\right)~,\qquad \Omega_\pm ^\psi ={1\over 2}\left(\Omega^R_{\pm}\mp \Omega^L_{\pm}\right)~.
\ee

The mass and angular momentum of the black hole are 
\be
M=\frac{(r_+^2+a^2)(r_+^2+b^2)}{2r_+^2}~,\qquad
J_\psi = {\pi\over 2}M\,b~,\qquad J_\phi = {\pi\over 2}M\,a~.
\ee 
The Bekenstein--Hawking entropy is
\be\label{SMP}
S= \frac{\pi^2(r_+^2+a^2)(r_+^2+b^2)}{2 r_+} =\pi^2 M r_+~.
\ee

There are two regular singular points at $r_\pm$ and an irregular singularity at infinity. Near the outer horizon $r\to r_+$ with
\be\label{eq:qwr}
R(x) \sim \left(x-{1\over2}\right)^{\pm i \alpha_+} \,,\qquad \alpha_+= {1\over 2\kappa_+}\left({\omega}-m_R {\Omega^R_+}-m_L {\Omega^L_+}\right) ~,
\ee
and near the inner horizon $r\to r_-$ with 
\be\label{eq:qwr2}
R(x) \sim \left(x+{1\over2}\right)^{\pm i\alpha_-} \,,\qquad \alpha_-= {1\over 2\kappa_-}\left({\omega}-m_R {\Omega^R_-}-m_L {\Omega^L_-}\right) ~ .
\ee
The identification of left- and right-moving sectors in terms of the monodromy coefficients proceeds almost exactly as in section \ref{sec:bhmono}.
The only subtlety is the presence of  two commuting $U(1)$ isometries,  $\partial_\phi$ and $\partial_\psi$, in addition to time translations $\partial_t$. This gives two inequivalent  ways of realizing the hidden conformal symmetry.
In the context of Kerr/CFT this was first discussed in \cite{Lu:2008jk,Hartman:2008pb}; see also \cite{Chen:2010ywa,Chen:2011kt,Chen:2011wm,Chen:2012ps,Chen:2012pt} and references therein.

 The central charge associated to 
  the $\phi$ circle is
\bea
c_\phi= 6 J_\psi~.
\eea
The boundary conditions used to derive this result are such that excitations along $\partial_\psi$ are frozen. In our language, this implies that the probe has  $m_\psi=0$.  Following the logic of section \ref{sec:HCS}, we identify left- and right-moving energies 
\be
 \omega_L := \alpha_+ - \alpha_-~,\qquad
 \omega_R:= \alpha_++ \alpha_-~,
\ee
with $m_\psi=0$. The identification $\phi\sim\phi +2\pi~$ then gives the temperatures 
\bea
T_{L,\phi}= {1\over 2\pi}{\kappa_- +\kappa_+\over \Omega_-^\phi-\Omega_+^\phi}={r_++r_-\over 2\pi b}
~,\qquad T_{R,\phi}={1\over 2\pi} {\kappa_-- \kappa_+\over \Omega_-^\phi-\Omega_+^\phi}={r_+-r_-\over 2\pi b}~,
\eea
in agreement with \cite{Krishnan:2010pv}.
Likewise, the central charge associated to the $\psi$ circle is
\bea
c_\psi= 6 J_\phi~,
\eea
which follows from boundary conditions that freeze the $\partial_\phi$ sector ($m_\phi=0$).  This gives
\bea
T_{L,\psi}= {1\over 2\pi}{\kappa_- +\kappa_+\over \Omega_-^\psi-\Omega_+^\psi}={r_++r_-\over 2\pi a}~,\qquad T_{R,\psi}={1\over 2\pi} {\kappa_-- \kappa_+\over \Omega_-^\psi-\Omega_+^\psi}={r_+-r_-\over 2\pi a}~.
\eea

It is easy to check that the Cardy formulas
\bea
S&=&\frac{\pi^2}{3}(c_\phi T_{L,\phi} + c_{\phi} T_{R,\phi})
=\frac{\pi^2}{3}(c_\psi T_{L,\psi} + c_{\psi} T_{R,\phi})~,
\eea
reproduce the black hole entropy \eqref{SMP}.

As with the Kerr black hole, this method allows us to identify the temperatures $T_{L,R}$ without taking a low frequency limit.
Moreover, the construction implemented here applies immediately to five-dimensional black rings. In this case, even though the wave equation is not separable, one can identify the monodromies associated to the inner/outer horizon. It is straight forward to check that the monodromies will be of the form \eqref{eq:qwr}--\eqref{eq:qwr2} and the rest of the analysis carries on. Further it agrees with the results in \cite{Castro:2012av,Chen:2012yd}.

\section{Kerr--AdS}\label{app:kads}

In this appendix, we derive monodromies for the four-dimensional Kerr--AdS geometry.  The metric is %\cite{Carter1968,Carter1968New}
\be\label{app:ads-geom}
ds^2={\Sigma\over {\Delta_r}}dr^2- {{\Delta_r}\over \Sigma}\left(dt -{a\over \Xi}\sin^2\theta\, d\phi\right)^2+ {\Sigma\over \Delta_\theta} d\theta^2+{   \Delta_\theta\over \Sigma}\sin^2\theta \left({(r^2+{a}^2)\over \Xi}\,d\phi-{{a}}\,dt\right)^2~,
\ee
where
\bea
\Delta_\theta&=&{1-{a^2\over \ell^2}\cos^2\theta}~, \quad \Xi=1-{a^2\over \ell^2}~\quad \Sigma=r^2+{a}^2\cos^2\theta~,
\cr
&&\Delta_r=(r^2 +{a}^2)(1+{r^2\over \ell^2})-2M r~=\prod_{i=1}^4(r-r_i)~.
\eea
$\Delta_r$ has four roots, denoted $r=r_i$, two of which are real and two of which are complex.  The largest real root is the outer horizon.  

The ADM mass and   angular momentum are
\be
M_{\rm ads}={M\over \Xi^2}~,\quad J_{\rm ads}={Ma\over \Xi^2}
\ee
Kerr--dS is obtained by taking $\ell^2\to -\ell^2$ (in this case, all roots of $\Delta_r$ are real for suitable ranges of the parameters). 

A massless field can be decomposed into modes 
\be
\psi = e^{-i\omega t+im\hat\phi}R(r)S(\theta)~,
\ee
where
\be
\hat \phi = \phi +{a\over \ell^2} t~.
\ee
%We will suppress the arguments $(\omega,m)$ from $R(r)$ and $S(\theta)$.  
The Klein--Gordon equation reduces to a spherical equation
\bea
\left[ {1\over \sin\theta}\partial_\theta
\left(\sin\theta\Delta_\theta\partial_\theta\right)-m^2{\Delta_\theta\over\sin^2\theta}+\omega^2 a^2\cos^2\theta {\Xi\over \Delta_\theta}\right] S(\theta)=-K_{\ell,{\rm ads}}  S(\theta)~,
\eea
and a radial equation
\bea\label{eq:MMM}
\Bigg[\partial_r \Delta_r \partial_r +\sum_i{(r_i^2 +a^2)^2\over \Delta_r'(r_i)}{(\omega-\Omega_i m)^2\over (r-r_i)}-\ell^2 \Xi \omega^2 +{a^2m^2\over \ell^2} \Bigg] R(r) =K_{\ell,{\rm ads}}  R(r)~ .
\eea
where 
\be
\Omega_i={a\over r_i^2+a^2}\left(1+{r_i^2\over \ell^2}\right)~.
\ee

Equation \eqref{eq:MMM} has five regular singular points located at $r=r_i$ and $r=\infty$.   We note that with asymptotic AdS boundary conditions, the singularity at $r=\infty$ is regular, rather than irregular---this is a generic feature of AdS boundary conditions. The monodromy eigenvalues at $r=r_i$ are
\be
\alpha_i= {(r_i^2 +a^2)\over \Delta_r'(r_i)}(\omega-\Omega_i m)~.
\ee
The singularity at $r=\infty$ is a resonant regular singularity, which means that the matrix $M_\infty$ has non-trivial Jordan block, so one solution has a logarithmic branch cut.  When the probe is instead massive, $r=\infty$ is a non-resonant regular singularity, so both solutions have algebraic branch cuts with monodromies directly related to the conformal dimension of the dual operator.

In either case, the monodromy matrices must obey the global identity
\be
M_1 M_2 M_3 M_4 M_\infty = \mathbf{1}~.
\ee
For each singular point we can define a set of modes, as in the case of Kerr.  In the present case, however,  with five singular points, it is not possible to obtain a generic expression for the scattering coefficients simply from monodromy data.

\bibliographystyle{utphys}
\bibliography{monodromies-bib}

\end{document}